\documentclass[11pt]{article}
\usepackage{latexsym}
\begin{document}
\thispagestyle{empty}
\begin{flushright}
MPI-PTh/95-117 \\
UNIGRZ-UTP-15-11-95 \\
November 1995 \\
hep-lat/9511019
\end{flushright}
\vskip 1cm
\begin{center}
{\huge{Lattice Schwinger Model with \vskip3mm
Interpolated Gauge Fields }}
\vskip10mm
\centerline{ {\bf
Christof Gattringer${}^*$ }}
\vskip 5mm
\centerline{Max-Planck-Institut f\"{u}r
Physik, Werner-Heisenberg-Institut}
\centerline{F\"ohringer Ring 6, 80805 Munich, Germany}
\vskip 2mm
\centerline{and}
\vskip 2mm
\centerline{Institut f\"{u}r Theoretische Physik der Universit\"at Graz}
\centerline{Universit\"atsplatz 5, 8010 Graz, Austria}
\vskip30mm
\end{center}
\begin{abstract}
\noindent
We analyze the Schwinger model on an infinite lattice using the continuum
definition of the fermion determinant and a linear interpolation of the
lattice gauge fields. The possible class
of interpolations for the gauge fields,
compatible with gauge invariance is discussed.
The effective action for the lattice gauge field
is computed for the Wilson formulation as well as for non-compact lattice
gauge fields. For the non-compact formulation we prove that the model has
a critical point with diverging correlation length at zero gauge coupling
$e$. We compute the chiral condensate for $e > 0$
and compare the result to
the N-flavor continuum Schwinger model. This indicates that there is only
one flavor of fermions with the same chiral properties as in the continuum
model, already before the continuum limit is performed. We discuss how
operators have to be renormalized in the
continuum limit to obtain the continuum Schwinger model.
\end{abstract}
\vskip20mm
\bigskip \nopagebreak \begin{flushleft} \rule{2 in}{0.03cm}
\\ {\footnotesize \
${}^*$ e-mail: chg@mppmu.mpg.de}
\end{flushleft}
\newpage
%
%-----------------------------------------------------------
%
\section{Introduction}
Although the study of lattice gauge theories with interpolated
gauge fields has a long history \cite{flume}-\cite{interpolnn},
the interest in this concept has considearbly increased
\cite{hsu}-\cite{kronfeld} recently. This increased interest is
due to 't Hooft's paper \cite{thooft} where it is argued
that interpolating the lattice gauge fields and using a carefully
regularized continuum fermion determinant provides an effective
lattice gauge theory carrying chiral fermions.
Some aspects of this approach have been discussed both in the Schwinger
model and the chiral Schwinger model
\cite{flume,gock,kronfeldal,montvay}.
However, no systematic treatment of the
(chiral) lattice Schwinger model with interpolated gauge fields and
continuum
prescription of the fermion determinant can be found in the literature.
This paper tries to close this gap for the standard Schwinger model with
vectorlike coupling.

The fact that the Schwinger model provides a testground for problems
involving chiral properties of fermions and species doubling is not the
only motivation for this analysis.
The model is one of the few instances where good analytic
control over a lattice theory with not too poor physical features can be
obtained. In particular the approach to the critical point and thus to the
continuum model can be analyzed.

The paper is organized as follows. We start with fixing the
action for the lattice gauge fields, where we discuss both the Wilson and
the non-compact formulation. Once the lattice action and thus the
corresponding lattice gauge transformation are fixed, they restrict the
class of possible interpolations. In particular a lattice gauge transformation
for the lattice gauge fields should give rise to a continuum gauge
transformation for the interpolated fields. This condition is sufficient
for the effective lattice action to be gauge invariant. From this
restriction, we derive a condition for the possible interpolations of
U(1)-lattice gauge fields. A linear interpolation which obeys this condition
is then used to compute the effective action. Both the Wilson and the
non-compact model can be quantized by integrating over the field strength
of the lattice fields, but only the latter gives rise to a Gaussian
measure and thus is easy to analyze. In what follows then we restrict
ourselves to the non-compact formulation. From the exponential falloff
of the two point function of the field strength we compute the correlation
length for small gauge coupling $e$. It turns out that the correlation
length diverges as $e$ approaches zero, giving rise to a critical point
at $e = 0$. In Section 5 we compute the chiral condensate for $e > 0$.
The result is compared to the chiral condensate of the continuum
Schwinger model with N flavors. This indicates that the lattice model
involves only one flavor degree of freedom with the same chiral properties
as in the continuum model already for finite correlation length. Finally we
discuss how operators have to be renormalized when performing
the continuum limit $e \rightarrow 0, et = $ const, where $t$ is the
distance in lattice units. We explicitely compute the wave function
renormalization constants for the field strength and the chiral densities.

%
%-----------------------------------------------------------
%

\section{Interpolation and gauge invariance}

The lattice under consideration is
$\mbox{Z\hspace{-1.35mm}Z}^2$, i.e. the lattice spacing is set to one.
Functions defined in the continuum can be identified by their
arguments $x,y \in \mbox{I\hspace{-0.7mm}R}^2$,
while lattice quantities have arguments
$n,m,k,r,s \in \mbox{Z\hspace{-1.35mm}Z}^2$.
We consider two types of actions for the U(1)-gauge fields.
The non-compact action
\begin{equation}
S_{NC} \; := \; \frac{1}{4} \sum_{{n} \in Z\mbox{\hspace{-1.35mm}}Z^2}
\; F_{ \mu \nu}({n}) F_{ \mu \nu}({n}) \; = \;
\frac{1}{2} \sum_{{n} \in Z\mbox{\hspace{-1.35mm}}Z^2}
\; F_{12}({n})^2 \; \;  ,
\end{equation}
with
\begin{equation}
F_{\mu \nu}({n}) \;  := \;  \Big( A_\nu({n} + \hat{e}_\mu) -
A_\nu({n}) - A_\mu({n} + \hat{e}_\nu) +
A_\mu({n}) \Big) \; .
\end{equation}
$A_\mu({n})$ may assume values in $(-\infty, + \infty)$ for all
${n} \in \mbox{Z\hspace{-1.35mm}Z}^2$
and $\mu = 1,2$. $\hat{e}_\mu$ is the unit
vector in $\mu$-direction.
We also consider the Wilson formulation of the gauge field action
\begin{equation}
S_{W} \; := \; \frac{1}{ e^2} \sum_{{n} \in Z\hspace{-1.35mm}Z^2}
\; \left[ 1 \; - \; \mbox{Re} \left( U_1({n}) U_2({n}\!+\!\hat{e}_1)
\overline{ U_1({n}\!+\!\hat{e}_2)}
\overline{ U_2({n})} \right) \right]
\; = \; \frac{1}{ e^2} \sum_{{n} \in Z\hspace{-1.35mm}Z^2}
\; \left[ 1 \; - \; \mbox{Re} \; e^{i e F_{12}({n})} \right] \;.
\end{equation}
In the last step the gauge transporters were expressed as
\begin{equation}
U_\mu({n}) \; := \; \exp\left( ie \; A_\mu({n}) \right) \; \; \;
\mbox{with} \; \; \; \; e A_\mu({n}) \; \in \; [ 0, 2\pi ] \; .
\end{equation}
Although we use the same symbol, it has to be kept in mind that the
$A_\mu({n})$ are restricted to the principal branch $[ 0, 2\pi/e]$
in the case of the Wilson action.

Both gauge field actions (1), (3) are invariant under the lattice
gauge transformation
\begin{equation}
A_\mu({n}) \; \longrightarrow \; A_\mu({n}) \; \; \; +
\; \Lambda({n} + \hat{e}_\mu) -
\Lambda({n})  \; ,
\end{equation}
where $\Lambda({n})$ is some scalar lattice field. The lattice
discretization of the derivatives in (1)-(5) could
have been chosen in a different
way, e.g. symmetric with respect to the sites. This would modify
the equation for the lattice gauge transformation and thus the class of
possible interpolations.

Having fixed the form of the gauge transformation of the lattice fields
we can start to think of an interpolation to the interior of the
lattice cells.
In order to study the class of possible interpolations,
we introduce the notation \cite{montvay} of an interpolation kernel
$D({x}; {n})$, and denote the interpolated gauge field as
($\mu$ not summed)
\begin{equation}
A_\mu({x}) \; := \; \sum_{{n} \in Z\hspace{-1.35mm}Z^2}
D_\mu ({x}; {n}) A_\mu({n}) \; \; \; \; \; \; \; {x} \in
\mbox{I\hspace{-0.7mm}R}^2 \; .
\end{equation}
We require the interpolation to be equivariant, i.e.~to transform the
lattice gauge transformation
(5) into a continuum gauge transformation
\begin{equation}
A_\mu({x}) \; \longrightarrow \;
A_\mu({x}) \; + \;
\frac{\partial}{\partial x _\mu} \Lambda({x}) \; ,
\end{equation}
for the interpolated vector field
$A_\mu({x})$,
which makes use of a continuum scalar field $\Lambda({x})$ obtained
from an interpolation of the lattice scalar $\Lambda({n})$
\begin{equation}
\Lambda({x}) \; := \; \sum_{{n} \in Z\hspace{-1.35mm}Z^2}
D^*({x}; {n}) \Lambda({n}) \; \; \; \; \; \; \; {x} \in
\mbox{I\hspace{-0.7mm}R}^2 \; .
\end{equation}
The interpolation kernel $D^*$
for the scalar field is not necessarily the same as the
kernel $D_\mu $ for the vector field.
The requirement that a lattice gauge transformation gives rise to a
continuum gauge transformation in the `interpolated world' is a sufficient
condition for the gauge invariance of the effective lattice model,
since a proper definition of
the continuum fermion determinant is invariant under continuum gauge
transformations. This requirement restricts the class of possible
interpolation kernels. Combining (5)-(8) gives rise to
\begin{equation}
\Big( D_\mu({x} ; {n} - \hat{e}_\mu) -
D_\mu({x} ; {n} ) \Big) \; \; \stackrel{{\bf !}}{=} \; \;
\frac{\partial}{\partial x _\mu} D^*({x} ; {n}) \; \; .
\end{equation}
This restriction which assumes the form (9) for U(1)-gauge fields also
in higher dimensionsform has to be obeyed by the interpolation kernels $D_\mu,
D^*$. A further restriction of possible solutions to (9) is that the
interpolation should be local in some sense; e.g.
$A_\mu({x}) = A_\mu({n})$ for ${x} =  {n}$. This is not a
physical requirement, but only a technical assumption in order to remain close
to the spirit of an interpolation.
Here we do not discuss the class of all possible interpolation kernels
that are solutions to (9), although this would be an interesting
challenge by itself. We interpolate the
fields as follows (here we denote the interpolation explicitely
instead of using the kernels $D_\mu$ and $D^*$ in order to simplify the
notation)
\begin{equation}
A_1({x}) \; := \; A_1({n}) \;  [1-t_2]  \;  + \;
A_1({n}+\hat{e}_2) t_2 \; \; \; \; , \; \; \; \;
A_2({x}) \; := \; A_2({n}) \;  [1-t_1]  \;  + \;
A_2({n}+\hat{e}_1) t_1 \; \; ,
\end{equation}
\begin{equation}
\Lambda({x}) \; := \; \Big( \Lambda({n}) [1-t_1] +
\Lambda({n}+\hat{e}_1) t_1 \Big) [1-t_2] \; + \;
\Big( \Lambda({n}+\hat{e}_2) [1-t_1] +
\Lambda({n}+\hat{e}_1+\hat{e}_2) t_1 \Big) t_2 \; \; ,
\end{equation}
for ${x} = {n} + {t} $ and $t_1,t_2 \in (0,1]$. It is
easy to check that the interpolation (10), (11) obeys the requirement of
transforming a lattice gauge transformation into a continuous one. The
interpolation (10) was already obtained \cite{flume, kronfeldal}
from another motivation, namely the
definition of a winding number for lattice gauge fields \cite{luscher}.
It has to be remarked, that the interpolation
of the gauge fields $A_\mu$ is not differentiable on the links
and in $\mu$-direction even not continuous. In particular the 1-component
$A_1$ is discontinuous with finite step height
on the links parallel to 2-direction and vice versa.
Thus the derivatives of the $A_\mu$ have to be computed in the sense of
distributions giving rise to $\delta$-distributions times the step height
at the lines of discontinuity.
However, things work out simpler if one considers gauge invariant
operators only, i.e. functions of $F_{12}({n})$.
Since $A_1$ is continuous in 2-direction and vice versa, the field strength
$F_{12}$ picks up no contributions involving $\delta$-distributions
\begin{equation}
F_{12}({x})
\; = \; \Big( A_2({n} + \hat{e}_1) -
A_2({n}) - A_1({n} + \hat{e}_2) +
A_1 ({n}) \Big) \; = \; F_{12}({n}) \; ,
\end{equation}
for ${x} = {n} + t $ and $t_1,t_2 \in (0,1]$.
The interpolated field strength is constant inside the lattice cells,
but discontinuous along the links.
For later use we quote the continuum Fourier transform
$\widetilde{F_{12}}({p})$
of the interpolated field strength given by (12)
\[
\widetilde{F_{12}}({p}) \;
:= \; \int_{-\infty}^{\infty} d^2 x \; F_{12} ({x})
e^{-i {p} \cdot {x}}
\]
\begin{equation}
 = \;
\sum_{{n} \in Z\hspace{-1.35mm}Z^2} \; F_{12}({n}) \;
e^{-i {p} \cdot {n}} \int_0^1 d^2t \; e^{-i{p} \cdot {t}}
\; = \; - \widehat{F_{12}}({p}) \; \frac{1-e^{-ip_1}}{ p_1} \;
\frac{1-e^{-ip_2}}{ p_2} \; ,
\end{equation}
where we introduced the lattice Fourier
transform $\widehat{F_{12}}({p})$ as
\begin{equation}
\widehat{F_{12}}(p) \; := \;
\sum_{{n} \in Z\hspace{-1.35mm}Z^2} \; F_{12}({n}) \;
e^{-i{p} \cdot {n}} \; .
\end{equation}
$\widehat{F_{12}}({p})$ is periodic in ${p}$ with respect to the
Brillouin zones.
%
%-----------------------------------------------------------------
%

\section{Effective lattice action and quantization}
The announced strategy is to quantize the fermions by a formal
Berezin path integral in the continuum \cite{berezin}.
Integrating out the continuum fermions gives rise to the fermion
determinant
$\mbox{det}[ \not\!\!{\partial} -
i e \not{\!\!\!A}]$.
The fermion determinant
is only defined when an ultraviolet and infrared
cutoff (for instance a finite space-time lattice also for the fermions,
\cite{weingarten})
is introduced. The determinant can then be normalized to 1 for $e = 0$, by
replacing it with $\mbox{det}[1 - K(A)]$ where
$K(A) = i e\not{\!\!\!\,A} {\not\!{\partial}}^{-1}$.
In two dimensions this determinant can be computed explicitly,
using the idea of regularized fermion determinants
(see e.g.~\cite{seiler}).
If we assume that the interpolated vector potential $A_\mu({x})$,
and thus the lattice gauge field $A_\mu({n})$ satisfies (see below)
some mild regularity and falloff conditions at infinity to make it
square integrable \cite{seiler}, the answer is
\begin{equation}
\mbox{det}_{reg}[ 1 - K(A) ] \; = \; \exp \left( -\frac{e^2}{2\pi}
\int_{-\infty}^{\infty} \frac{ d^2 p}{(2\pi)^2} \;
\widetilde{F_{12}}(- {p}) \;
\frac{1}{{p}^{2}} \; \widetilde{F_{12}}({p}) \right) \; .
\end{equation}
It has to be remarked, that the falloff
condition corresponds to zero topological charge.
Expression (15) makes sense only if
$\widetilde{F_{12}}({p})$ vanishes at zero
momentum which in turn requires $0 =
\widehat{F_{12}}({0}) =
\sum_{{n}} [ A_2({n} + \hat{e}_1) -
A_2({n}) - A_1({n} + \hat{e}_2 ] +
A_1({n}) \big)$. The last equation is always fulfilled if the
lattice gauge fields $A_\mu({n})$ are absolutely summable over
Z\hspace{-1.35mm}Z$^2$. This is the announced falloff condition expressed
in terms of the lattice gauge fields. It can e.g. be imposed by
restricting the $A_\mu({n})$ to a finite rectangle in
Z\hspace{-1.35mm}Z$^2$.
Inserting (13) one obtains for the
contribution of the fermion determinant to the effective gauge field action
\begin{equation}
S_F \; := \; \frac{e^2}{2\pi}
\int_{-\pi}^{\pi}
\frac{d^2p}{(2\pi)^2} \; \widehat{F_{12}}(-{p})
\widehat{F_{12}}({p}) \sum_{{k} \in Z\hspace{-1.35mm}Z^2}
\frac{1}{({p} + 2\pi {k}\,)^2} \;
\frac{2-2\cos( p_1)}{(p_1 +  2 \pi k_1)^2} \;
\frac{2-2\cos( p_2)}{(p_2 +  2 \pi k_2)^2} \; ,
\end{equation}
where the periodicity of $\widehat{F_{12}}({p})$
was used, in order to restrict the integration to the first Brillouin zone.

Fourier transforming the non-compact gauge field action (1)
and adding it to (16) gives the effective lattice action
\begin{equation}
S_{EFF} :=
\frac{1}{2}
\int_{-\pi}^{\pi} \frac{d^2p}{(2\pi)^2} \; \widehat{F_{12}}(-{p}\,)
\widehat{F_{12}}({p}\,) \left[1+
\frac{e^2}{\pi}\sum_{{k} \in Z\hspace{-1.35mm}Z^2}
\frac{1}{({p} + 2\pi {k})^2} \;
\frac{2\!-\!2\cos(p_1)}{(p_1 +  2 \pi k_1)^2} \;
\frac{2\!-\!2\cos(p_2)}{(p_2 +  2 \pi k_2)^2} \right] \; .
\end{equation}
The effective action for non-compact gauge fields is a quadratic form
in the field strength $F_{12}$.
As in the continuum, when restricting to gauge
invariant observables, the model can be quantized by defining a path integral
for $F_{12}$ (see e.g.~\cite{seiler}) and no extra gauge fixing term has to
be taken into account.
As it stands, the effective action (17) is restricted to
$\widehat{F_{12}}({p})$ which are $L^2$ in the first Brillouin zone
and vanish at zero momentum. In order to define a proper Gaussian measure for
$F_{12}$, the latter requirement has to be abolished by the introduction
of a cutoff.
However as in the continuum the resulting measure converges in the
weak sense (i.e.~moments and characteristic function converge) to a
Gaussian measure $d\mu_C[F_{12}]$ with lattice-covariance
\begin{equation}
\widehat{C}({p} ) \; := \;
\left[ \; 1 \; + \; \frac{e^2}{\pi}\sum_{{k} \in Z\hspace{-1.35mm}Z^2}
\frac{1}{({p} + 2\pi {k})^2} \;
\frac{2-2\cos(p_1)}{(p_1 +  2 \pi k_1)^2} \;
\frac{2-2\cos(p_2)}{(p_2 +  2 \pi k_2)^2} \right]^{-1} \; .
\end{equation}
The Gaussian measure $d\mu_C[F_{12}]$ with covariance
$C$ is the starting point for quantizing the effective lattice gauge theory
in the non-compact formulation.

A measure for the model using the Wilson formulation (3) can
be defined by integrating over the link variables $U_\mu(n)$.
In two dimensions for U(1) gauge fields
this can be done by integrating each $F_{12}({n})$ over
$[ -\pi/e, \pi/e ]$. However, the resulting measure is not Gaussian
and thus much more involved. In the following we will restrict ourselves
to the non-compact formulation.

%
%-----------------------------------------------------------------
%

\section{The correlation length for small gauge coupling}
In order to compute the correlation length of the effective lattice
gauge theory, we evaluate the two point function of the field strength.
It will be shown that for small $e$ the two point function falls
off exponentially for large time-like separation, thus defining a
correlation length. The two point function of $F_{12}({n})$ is simply
given by the inverse Fourier transform of the covariance (18)
\[
\langle F_{12}(0,t) \; F_{12}(0,0) \rangle \; = \;
\int_{-\pi}^{\pi} \frac{d^2p}{(2\pi)^2} \; e^{-ip_2t} \;
\widehat{C}({p})
\; = \; \int_{-\pi}^{\pi} \frac{d^2p}{(2\pi)^2} \; e^{-ip_2t}
\left[ 1 - \frac{e^2/\pi}{\sigma(p_1,p_2 )^{-1} +
\frac{e^2}{\pi} } \right]
\]
\begin{equation}
= \; \delta_{0,0} \delta_{0,t} \; - \; \frac{e^2}{\pi}
\int_{-\pi}^{\pi} \frac{d^2p}{(2\pi)^2} \; e^{-ip_2t}
\frac{1}{\sigma(p_1,p_2)^{-1} +
\frac{e^2}{\pi} } \; .
\end{equation}
In the last expression the contact term which is also known from
the continuum
was split of. We furthermore introduced the abbreviation
\begin{equation}
\sigma(p_1,p_2) \; := \; \sum_{{k} \in Z\hspace{-1.35mm}Z^2}
\frac{1}{({p} + 2\pi {k})^2} \;
\frac{2\!-\!2\cos(p_1)}{(p_1 +  2 \pi k_1)} \;
\frac{2\!-\!2\cos(p_2)}{(p_2 +  2 \pi k_2)} \; .
\end{equation}
In order to extract the exponential falloff of the two point
function, the singularities of
\begin{equation}
f(p_1,p_2) \; := \; \frac{1}{\sigma(p_1,p_2)^{-1} +
\frac{e^2}{\pi} } \;
\end{equation}
in the complex $p_2$-plane have to be computed.
For small $|{p}|$, $\sigma(p_1,p_2)^{-1}$ behaves as $p_1^2 + p_2^2$,
(as should be for any proper dispersion function). Thus one indeed can
expect to find a mass gap. Unfortunately, due to the
form of $\sigma(p_1,p_2)$, the explicit computation of the singularities
requires the solution of transcendental equations. However for small
gauge coupling $e$, the pole structure can be analyzed perturbatively.
$\sigma(p_1,p_2)$ has the following symmetry properties
\[
\sigma(-p_1,p_2) \; = \; \sigma(p_1,p_2) \; = \; \sigma(p_1,-p_2) \; ,
\]
\[
\sigma(p_1\!+\!2\pi j,p2) \; = \;
\sigma(p_1,p_2) \; = \; \sigma(p_1,p_2\!+\!2\pi j)
\; \; \; \; \; \;
\forall \; j \; \in \; \mbox{Z\hspace{-1.35mm}Z} \; ,
\]
\begin{equation}
\sigma(p1,\overline{p_2}) \; = \; \overline{\sigma(p_1,p_2)} \; ,
\end{equation}
where $p_2$ is now a complex variable. As can be seen
from (21) those symmetry properties carry over to $f(p_1,p_2)$. Since
$\big( 2-2\cos(z) \big)/z^2$ is an analytic function, the only singularities
in $\sigma(p_1,p_2)$ come from the $\big({p} + 2\pi {k}\big)^{-2}$
terms in (20). This implies that the singularities
of $\sigma(p_1,p_2)$ in the
complex $p_2$-plane are situated at
\begin{equation}
p_2 \; := \; - 2 \pi j \; \pm \; i( p_1 + 2 \pi l) \; \; \; \; \; \; \; \;
j,l \in \mbox{Z\hspace{-1.35mm}Z} \; .
\end{equation}
$\sigma(p_1,p_2)^{-1}$ for fixed $p_1$ is a meromorphic function
in the complex $p_2$-plane with zeroes given also by (23).
Thus for small $e$, the poles of $f(p_1,p_2)$ lie in the vicinity of
the $p_2$ given by (23).
We expand $\sigma(p_1,p_2)^{-1}$ around the zero $p_2 = ip_1$ by setting
\begin{equation}
p_2 \; := \; ip_1 \; + \; \varepsilon \; + \; i \delta \; \; \; \;
\mbox{with} \; \; \; \varepsilon, \delta \in \mbox{I\hspace{-0.7mm}R},
\; \; |\varepsilon|, |\delta| \ll 1 \; .
\end{equation}
Due to the symmetry properties (22)
of $\sigma(p_1,p_2)$ it is sufficient to study the expansion around the
zero $p_2 = ip_1$. All the other zeroes can be obtained by applying
the symmetry transformations (22). The integrand $f(p_1,p_2)$ can
be rewritten identically to (using (24))
\[
f(p_1,p_2) = \left( \frac{2\!-\!2\cos(p_1)}{p_1^2}
\frac{2\!-\!2\cos(ip_1\!+\!\varepsilon\!+\!i\delta)}
{(ip_1\!+\!\varepsilon\!+\!i\delta)^2}
\;+\;\Big[ \varepsilon^2\!-\!\delta^2\!-\!2 p_1\delta\!+\!
i 2 \varepsilon( p_1 \!+ \!\delta) \Big]
\sigma^\prime(p_1, ip_1\!+\!\varepsilon\!+\!i\delta) \right)
\]
\[
\times \;
\Bigg(
\varepsilon^2 \; - \; \delta^2 \; - \;
2 p_1 \delta \; + \; i 2 \varepsilon(p_1 + \delta)
\]
\begin{equation}
+\; \frac{e^2}{\pi} \left[ \frac{2\!-\!2\cos(p_1)}{p_1^2}
\frac{2\!-\!2\cos(ip_1\!+\!\varepsilon\!+\!i\delta)}
{(ip_1\!+\!\varepsilon\!+\!i\delta)^2}
+ \Big[ \varepsilon^2\!-\!\delta^2\!-\!2 p_1\delta\!+\!
i 2 \varepsilon( p_1\!+\!\delta) \Big]
\sigma^\prime(p_1, ip_1\!+\!\varepsilon\!+\!i\delta) \right] \Bigg)^{-1} \; ,
\end{equation}
where $\sigma^\prime(p_1,p_2)$ is obtained from $\sigma(p_1,p_2)$
by omitting the ${k} = (0,0)$ term in the sum (20).
$\sigma^\prime(p_1,p_2)$ has no more singularities in the first Brillouin
zone.
Now we restrict ourselves to small $e$, in particular we assume
\begin{equation}
e \; \ll \; 1 \; \; \; \; \mbox{and}
\; \; \; \; \varepsilon, \delta \; \sim \; O(e) \; .
\end{equation}
It has to be remarked, that we consider small but fixed $e$ and extract
the exponential falloff for $t \rightarrow \infty$. Later the joint limit
$e \rightarrow 0 \; , \; et = $ const which leads to the continuum
Schwinger model will be discussed.
The denominator of $f(p_1,p_2)$ is given by
\begin{equation}
\varepsilon^2 - \delta^2 - 2 p_1\delta\ \; + \; \frac{e^2}{\pi}
\frac{2\!-\!2\cos(p_1)}{p_1^2} \; \frac{2\cosh(p_1)-2}{p_1^2}
\; \; + \; \; i \; 2 \varepsilon(p_1 + \delta) \; \; + \; \;
O(e^3) \; .
\end{equation}
To compute the pole up to order $O(e^2)$ the
denominator (27) has to vanish up to this order. This gives two
equations for the real and the imaginary part, which are used
to compute $\varepsilon$ and $\delta$. The equation for the
imaginary part has the two solutions $\epsilon = 0$ ($\delta$
arbitrary) and
$\delta = - p_1$ ($\varepsilon$ arbitrary).
When inserting the latter into the equation for
the real part, it does not allow for a real $\varepsilon$ and thus is
ruled out. Inserting the other solution $\varepsilon = 0$ into the
equation for the real part gives
\begin{equation}
\delta^2 \; + \; 2 p_1\delta\ \; - \; \frac{e^2}{\pi}
\frac{2 - 2\cos(p_1)}{p_1^2} \; \frac{2\cosh(p_1)-2}{p_1^2}
\; = \; 0 \; .
\end{equation}
In case that $p_1 \sim O(e)$ this is an equation of
$O(e^2)$ only, in case of $p_1 \sim 1$ it is an
equation of $O(e)$ with all
corrections up to $O(e^2)$ included. Anyway we will
solve (28) as a quadratic equation for $\delta$ and thus obtain the
correct result up to $O(e^2)$. This gives rise to
\begin{equation}
\delta \; = \; - p_1 \; \pm \; \beta(p_1) \; ,
\end{equation}
where we introduced the abbreviation
\begin{equation}
\beta(p_1) \; = \;
\sqrt{p_1^2 + \frac{e^2}{\pi}
\frac{2 - 2\cos(p_1)}{p_1^2} \; \frac{2\cosh(p_1)-2}{p_1^2}} \; .
\end{equation}
The solutions with the minus sign
in front of the square root is related to the
plus-solution by the transformations (22). Inserting $\varepsilon = 0$,
and the solutions (29) for $\delta$ into (24), and applying
the transformations (22) to the result gives the positions of
all poles up to $O(e^2)$
\begin{equation}
p_2 \; = \; \Big[ 2\pi j \; \pm \; i\Big( 2\pi l +
\beta(p_1) \Big) \Big] \; \Big( 1 + O(e) \Big) \; \; \;
\; \; \; \; \; j,l \in \mbox{Z\hspace{-1.35mm}Z} \; .
\end{equation}
Note that $p_1$ is restricted to $[-\pi,\pi]$.
The exponential decay of the two point function is determined by
the pole with $j,l = 0$ and the plus sign in (31), which is given by
$p_2 = i \beta (p_1) [ 1 + O(e) ] := p_2^0$.
In order to apply the
residue theorem to the complex $p_2$-integration we also need the residue
at $p_2^0$. A simple computation gives
\begin{equation}
\mbox{Res}_{p_2 = p_2^0 } \; \;e^{i t p_2 }f(p_1,p_2) \; = \;
\frac{-i\;e^{-t \beta(p_1)\; [1+O(e)]} } \;
{2 \beta(p_1)} \frac{2\!-\!2\cos(p_1)}{p_1^2} \;
\frac{2\cosh(\beta(p_1))\!-\!2}{\beta(p_1)^2} \;
\Big(1 + O(e) \Big).
\end{equation}
Now the $p_2$ integration can be solved easily by applying the residue
theorem to a contour integral in the complex $p_2$-plane. The contour is
a rectangle in the upper half plane with a piece
from $-\pi$ to $\pi$ along the real axis, and a piece
from $\pi + iR$ to $-\pi+iR$ parallel to the real axis, where
R is some large, real and positive number. These two pieces are connected
by two straight lines parallel to the imaginary axis. Due to the symmetry
properties (22) the contributions along those two lines cancel
each other. Furthermore the contribution from the piece parallel to the
real axis decreases exponentially for large $R$. Thus the integral along
the real axis from $-\pi$ to $\pi$ is $i 2 \pi$ times
the sum over the residues of all poles inside the contour. The exponential
falloff is dominated by the pole $p_2^0$. The contributions from the
other poles decay exponentially stronger
and their residues can be summed. One obtains
\[
\langle F_{12}(0,t) \; F_{12}(0,0) \rangle \; = \;
\delta_{0,0} \delta_{0,t} \; - \; \frac{e^2}{\pi}
\int_{-\pi}^{\pi} \frac{d^2p}{(2\pi)^2} \; e^{-ip_2t}
f(p_1,p_2)
\]
\begin{equation}
= \; \delta_{0,0} \delta_{0,t} \; - \; \frac{e^2}{\pi}
\int_{-\pi}^{\pi} \frac{dp_1}{\pi} \;
\frac{e^{ -t \beta(p_1) [ 1 + O(e) ] }}
{\beta(p_1)} \frac{2\!-\!2\cos(p_1)}{p_1^2} \;
\frac{2\cosh(\beta(p_1))\!-\!2}{\beta(p_1)^2} \;
\Big( 1 + O(e)
+ O(e^{-t \pi}) \Big) \; .
\end{equation}
The integrand of the remaining $p_1$-integration has no more
singularities. Since $\beta(p_1)$ is bounded from below by $e/\sqrt{\pi}$,
one concludes
\begin{equation}
\Big| \langle F_{12}(0,t) \; F_{12}(0,0) \rangle \Big| \; \leq \;
\frac{e^2}{\pi} \; C(e) \;
\exp \Big( -\frac{t}{\xi}[1 + O(e)] \Big) \; ,
\end{equation}
where the correlation length $\xi$ is given by
\begin{equation}
\xi \; = \; \frac{\sqrt{\pi}}{e} \; .
\end{equation}
Thus we have established exponential decay of the two-point
function of $F_{12}$.
The correlation length diverges at $e = 0$ giving rise
to a critical point, where the continuum limit can be taken.

Beside the factor $e^2$ on the right hand side
of (34) a factor $C(e)$, constant in $n$ shows up,
which behaves as $\ln(e)$ for small $e$.
However, in Section 6 we will show that when performing the
continuum limit $e \rightarrow 0 \; , \; t/\xi = $ const, no
such extra logarithmic divergence emerges.

%
%-----------------------------------------------------------------
%

\section{Chiral condensate}

The chiral condensate is a rather instructive expectation value in the
Schwinger model. In particular for the N-flavor continuum Schwinger model one
obtains \cite{belvedere, seilergatt}
\begin{equation}
\lim_{|{x} - {y}| \rightarrow \infty} \Big\langle
\prod_{a=1}^j \overline{\psi}^{(a)}(x) P_+ \psi^{(a)}(x) \;
\overline{\psi}^{(a)}(y) P_- \psi^{(a)}(y) \Big\rangle \; = \;
\left\{ \begin{array}{cc}
\Big(-\frac{e_c^2N}{16\pi^3}e^{2\gamma} \Big)^N & \mbox{for}
\; j = N \\ 0 & \mbox{for} \; j < N
\end{array} \right. \; ,
\end{equation}
where $a$ is the flavor index, $P_\pm$ denote the projectors on left and
right handed chirality, $\gamma$ is Euler's constant and finally $e_c$ denotes
the gauge coupling of the continuum model, where we introduced the subscript
$c$ for conceptual hygene. The chiral condensate of the N-flavor model
involves chiral densities of all flavors, and thus is sensitive to the number
of flavors. One can learn about the doubling problem and the chiral
properties of the lattice
model by comparing it's condensate to the condensate of the N-flavor
continuum model. On the lattice we will evaluate
\begin{equation}
\langle \overline{\psi}({n}) P_+ \psi({n}) \;
\overline{\psi}({m}) P_- \psi({m}) \rangle \; = \;
- \int d \mu_C [ F_{12}] \; G_{12}({n},{m};A_\mu)
G_{21}({m},{n};A_\mu) \; .
\end{equation}
$G$ is the continuum fermion propagator in an external field given by
\cite{schwinger}
\begin{equation}
G({n},{m};A_\mu) \; = \;  \frac{1}{2\pi} \frac{\gamma_\mu
(n_\mu - m_\mu)}{({n} - {m})^2} \;
e^{i[\Phi({n}) - \Phi({m})]}  \; ,
\end{equation}
with
\begin{equation}
\Phi({n}) \; = \; -\int d^2x  D({n}-{x})
\Big( \partial_\mu A_\mu(x) + i \gamma_5
\varepsilon_{\mu \nu} \partial_\mu A_\nu({x}) \Big) \; =: \;
\theta({n}) + i \gamma_5 \chi({n}) \; ,
\end{equation}
and $D$ denotes the Green's function of $-\triangle$.
In the last step we introduced the longitudinal part
$\theta = \triangle^{-1} \partial_\mu A_\mu$ which cancels in
gauge invariant expectation values like (37), and the gauge invariant
part
\begin{equation}
\chi({n}) \; := \; \triangle^{-1} F_{12}({n}) =
\int_{-\pi}^{\pi} \frac{d^2p}{(2\pi)^2} \; \widehat{F_{12}}({p}\,)
e^{i{p}\cdot{n}}
\sum_{{k} \in Z\hspace{-1.35mm}Z^2} \;
\frac{1}{({p}+2\pi{k})^2} \;
\frac{1-e^{-ip_1}}{ p_1 + 2\pi k_1} \;
\frac{1-e^{-ip_2}}{ p_2 + 2\pi k_2} \; ,
\end{equation}
where the Fourier transform (13) of the interpolated $F_{12}$ was inserted.
The Gaussian integral in (37) can be solved, giving rise to
(for simplicity we choose the space-time arguments to be ${n} = (0,t),
{m} = (0,0)$)
\begin{equation}
\langle \overline{\psi}(0,t) P_+ \psi(0,t) \;
\overline{\psi}(0,0) P_-\psi(0,0) \rangle \; = \;
- \; \frac{1}{(2\pi)^2} \; \frac{1}{t^2}\; e^{2E(t,e)} \; ,
\end{equation}
where we defined
\begin{equation}
E(t,e) \; = \; e^2
\int_{-\pi}^{\pi} \frac{d^2p}{(2\pi)^2} \; [2-2\cos(p_2t)] \;
\frac{\rho(p_1,p_2)}{1 + \frac{e^2}{\pi} \sigma(p_1,p_2)} \; .
\end{equation}
$\rho(p_1,p_2)$ is given by
\begin{equation}
\rho(p_1,p_2) \; := \; \sum_{{r},{s} \in Z\hspace{-1.35mm}Z^2} \;
\frac{1}{({p}+2\pi{r})^2} \;
\frac{1}{({p}+2\pi{s})^2} \;
\frac{2-2\cos(p_1)}{ (p_1 + 2\pi r_1)(p_1 + 2\pi s_1)} \;
\frac{2-2\cos(p_2)}{ (p_2 + 2\pi r_2)(p_2 + 2\pi s_2)} \; .
\end{equation}
The condensate is now being formed by $E(t,e)$ which contains a
term proportional to $\ln(t)$ which cancels the $1/t^2$ factor in (41),
plus a term which approaches a finite constant for $t \rightarrow \infty$.
Thus we have to split off the logarithm. The integrand in (42) can be
rewritten identically to give
\begin{equation}
E(t,e) \; = \; \frac{1}{2\pi} \int_{-\pi}^{\pi} d^2 p \;
[1-\cos(p_2 t)] \; \frac{1}{p_1^2 + p_2^2} \; - \;
\frac{1}{2\pi} \int_{-\pi}^{\pi} d^2 p \;
[1-\cos(p_2 t)] \; I({p}) \; ,
\end{equation}
with
\begin{equation}
I({p}) \; := \;
\frac{1 + \frac{e^2}{\pi} [ \sigma^\prime(p_1,p_2) - {p}^{2} \;
\rho^\prime(p_1,p_2) ]}{\frac{e^2}{\pi} \frac{2-2\cos(p_1)}{p_1^2}
\frac{2-2\cos(p_2)}{p_2^2} + {p}^{2} [ 1 +
\frac{e^2}{\pi} \sigma^\prime(p_1,p_2)]}
\end{equation}
$\rho^\prime(p_1,p_2)$ is obtained from $\rho(p_1,p_2)$ by omitting the
term with ${r}={s}=(0,0)$ in the sum (43).
${p}^{2} \; \rho^\prime(p_1,p_2)$ and $\sigma^\prime(p_1,p_2)$
have no more poles in the first Brillouin zone.
By applying the Riemann-Lebesgue
Lemma, the second integral can easily be shown to
approach a finite constant for $t \rightarrow \infty$. Thus one is left
with the problem of extracting the logarithm from the first integral in (44).
This integral can again be treated by applying the residue theorem in the
complex $p_2$-plane. When using the same contour as for the correlation length,
the contribution along $x+iR, x \in [-\pi, \pi]$
falls off exponentially with $R$. The contributions
along the lines parallel to the imaginary axis do not cancel each other
this time,
but give rise to the term ($R \rightarrow \infty$)
\begin{equation}
R(t) \; := \;  2 \int_{-\pi}^{\pi} dp_1 \int_0^\infty
dy \; \frac{y [ 1 - (-1)^t \; e^{-ty} ]}{(p_1^2 + \pi^2 - y^2)^2 +
4\pi^2 y^2} \; ,
\end{equation}
which aproaches a constant for $t \rightarrow \infty$.
Thus when applying the residue theorem to the first term in (44)
one obtains
\begin{equation}
\frac{1}{2\pi} \int_{-\pi}^{\pi} d^2 p \;
[1-\cos(p_2 t)] \; \frac{1}{p_1^2 + p_2^2} \; = \;
\int_0^\pi dp_1 \frac{1-e^{-tp_1}}{p_1} \; + \; R(t) \; .
\end{equation}
The logarithm from the $p_1$ integral can be made explicit
by using the exponential integral (see e.g. \cite{oberhettinger}
for the expansion)
\begin{equation}
\mbox{E}_1(x) \; = \;
- \mbox{Ei}(-x) \; = \; \int_x^\infty d\tau \; \frac{e^{-\tau}}{\tau}  \; =
\; - \; \gamma -\ln(x) - \sum_{l=1}^\infty \frac{(-x)^l}{l \cdot l!} \; .
\end{equation}
One obtains
\begin{equation}
\lim_{\varepsilon \rightarrow 0}
\int_\varepsilon^\pi dp_1 \frac{1-e^{-t p_1}}{p_1} \; = \;
\lim_{\varepsilon \rightarrow 0} \Big( \ln(\pi) - \ln(\varepsilon)
-\mbox{E}_1(\varepsilon t) + \mbox{E}_1(\pi t) \Big) \; = \;
\ln(t) + \gamma + \ln(\pi) + \mbox{E}_1(\pi t) \; ,
\end{equation}
where $\gamma$ denotes the Euler constant.
Putting things together, the exponent $E(t,e)$ reads
\begin{equation}
E(t,e) \; = \; \ln(t) + R(t,e) \; ,
\end{equation}
where $R(t,e) := R(t) + \gamma + \ln(\pi) + \mbox{E}_1(\pi t) + E_2(t,e)$
collects all terms, which were already shown to approach
a finite ($e$-dependent) constant,
giving $\lim_{t \rightarrow \infty} R(t,e):= r(e)$.
We end up with the following result for the condensate at $e > 0$
\begin{equation}
\lim_{t \rightarrow \infty}
\langle \overline{\psi}(0,t) P_+ \psi(0,t) \;
\overline{\psi}(0,0) P_-\psi(0,0) \rangle\; = \;
- \; \frac{1}{(2\pi)^2} \; e^{2 r(e)} \; \neq \; 0 \; .
\end{equation}
The result (51) is exact for all $e>0$ and contains no expansion
in $e$. The limit $e \rightarrow 0$ will be discussed in the next
section.

The nonvanishing condensate for finite $e$ is an important result.
If there were doublers in the model under consideration, the chiral
condensate would include higher powers of the chiral densities,
and the right hand side of (51) would be zero.
However since we used the continuum determinant no doublers can be expected
and we conclude that the formation of the chiral condensate
at $e > 0$ is a strong indication that there is only one flavor of
fermions which has the chiral properties of the continuum
already for finite correlation length. Also the continuum limit discussed
in the next section supports this result.

%
%-----------------------------------------------------------
%

\section{Continuum limit}
In Section 4 it was shown (compare Eq. (35))
that the correlation length is given by
$\xi = \sqrt{\pi}/e$, where $e$ is the coupling of the lattice model.
We now define our length scale $L_0$ to be proportional to the correlation
langth, i.e~$L_0 := \lambda \xi$. A physical distance $|x|$ is
measured in units of $L_0$ giving rise to $|x| = t/L_0$.
The continuum gauge coupling $e_c$ which has the dimension
of a mass is defined as $e_c = eL_0$. Thus we obtain
for the ratio $t/\xi$
\begin{equation}
\frac{t}{\xi} \; = \; t \frac{e}{\sqrt{\pi}} \; = \;
\mbox{const} \; = \; |{x}| \frac{e_c}{\sqrt{\pi}} \; .
\end{equation}
Since (34) holds for all $t$ (and small $e$), the prescription (52)
reproduces the exponential decay of the corresponding two-point function
in the continuum (K$_0$ denotes the modified Bessel function)
\begin{equation}
\langle F_{12}({x}) \; F_{12}(0) \rangle \; = \;
\delta^{(2)}({x}) \; - \; \frac{e_c^2}{\pi} \frac{1}{2\pi} \;
\mbox{K}_0\Big( \frac{e_c}{\sqrt{\pi}} | {x}| \Big) \; \; \;
\stackrel{ | {x}| \gg 1}{\sim} \; \; \;
\; e^{-\frac{e_c}{\sqrt{\pi}} | {x}|} \; .
\end{equation}
When sending $e \rightarrow \infty, et =$ const, in (34) the correct
overall factor depending on $e$ has to be determined and included
into a wave function renormalization $Z_{12}(e)$ for the field strength
At the end of Section 4 it was noticed, that the extra factor
$C(e) \sim \ln(e)$
on the right hand side of (34) is not there when one considers the
joint limit $e \rightarrow \infty, et =$ const. To establish this one
has to prove that (compare (19))
\begin{equation}
I(t,e) \; := \; \int_{-\pi}^\pi \frac{d^2p}{(2\pi)^2} \;
e^{i p_2 t} \; \frac{1}{\sigma(p_1,p_2)^{-1} + \frac{e^2}{\pi}} \; \;
\stackrel{ {e \rightarrow 0 \atop et = const}}{
-\!\!\!-\!\!\!-\!\!\!-\!\!\!-\!\!\!-\!\!\!-\!\!\!-\!\!\!\!\!\longrightarrow}
\; \; \mbox{const} \; < \; \infty \; .
\end{equation}
Define
\begin{equation}
J(t,e) \; := \; \int_{-\pi}^\pi \frac{d^2p}{(2\pi)^2} \;
e^{i p_2 t} \; \frac{1}{{p}^{2} + \frac{e^2}{\pi}} \; .
\end{equation}
This gives rise to
\begin{equation}
f(t,e) \; := \; I(t,e) - J(t,e) \; = \;
\int_{-\pi}^\pi \frac{d^2p}{(2\pi)^2} \;
e^{i p_2 t} \; \frac{ \frac{2-2\cos(p_1)}{p_1^2}\frac{2-2\cos(p_2)}{p_2^2}
-1 + {p}^{2} \sigma^\prime(p_1,p_2)}{
[1 + \frac{e^2}{\pi}\sigma(p_1,p_2)][p^{2} + \frac{e^2}{\pi}]} \; .
\end{equation}
The integrand in (56) has no more infrared singularities, and the limit
$e \rightarrow 0$ can be performed without letting $t \rightarrow \infty$
at the same time. Furthermore for all values of $e$ the limit
$t \rightarrow \infty$ gives zero due to the Riemann-Lebesgue lemma.
Thus $f(t,e)$ approaches zero when taking the joint limit in the
sense of (52). Finally performing the change of variables
${q} := {p} \sqrt{\pi}/ e$ and using (52) one obtains
\begin{equation}
J(t,e) \; := \; \int_{-\frac{\pi^{3/2}}{e}}^{\frac{\pi^{3/2}}{e}}
\frac{d^2q}{(2\pi)^2} \;
e^{i q_2 \frac{e}{\sqrt{\pi}}t} \; \frac{1}{{p}^{2} + 1} \; = \;
\frac{1}{2\pi} \mbox{K}_0 \Big(\frac{e_c}{\sqrt{\pi}} |{x}\,| \Big)
\; + \; o(e) \; .
\end{equation}
Comparing (57) and
(53) we find that we have obtained much more than simply the correct power
of $e$. Indeed we reproduced the functional form of the continuum
two point function of the field strength. Defining the
wave function renormalization constant as
\begin{equation}
Z_{12}(e) \; := \; \frac{e_c^2}{e^2} \; ,
\end{equation}
one ends up with (use (19), (56) and (57))
\begin{equation}
\lim_{e \rightarrow 0 \atop et = const} \; Z_{12}(e)
\langle F_{12} (0,t) \; F_{12} (0,0) \rangle \; = \;
\delta^{(2)}({x}) \; - \; \frac{e_c^2}{\pi} \frac{1}{2 \pi}
\mbox{K}_0 \Big(\frac{e_c}{\sqrt{\pi}} |{x}| \Big) \; ,
\end{equation}
where even the contact term of the continuum is reproduced since
$\delta_{0,t}\delta_{0,0} e_c^2/e^2 \rightarrow  \delta^{(2)}({x})$.

When performing the continuum limit for the two point function of
the chiral densities, we did not get that far, since in (42) the
Riemann-Lebesgue type argument cannot be applied
(the exponent (42) is not a
Fourier transform), which was used to obtain
the functional form of the two point function of the field strength.
However the necessary wave
function renormalization for the chiral densities
can be computed explicitely. In particular we
will show that the exponent $E(t,e)$ in (41) remains bounded in the continuum
limit. It can be rewritten to $E(t,e) = E_1(t,e) + E_2(t,e)$ where
\begin{equation}
E_1(t,e) \; := \;
\int_{-\pi}^\pi \frac{d^2p}{(2\pi)^2} \;
\frac{2-2\cos(p_2 t)}{{p}^{2}} \;
\frac{ \frac{e^2}{{p}^{2}}
\frac{2-2\cos(p_1)}{p_1^2}\frac{2-2\cos(p_2)}{p_2^2}}{
1 + \frac{e^2}{\pi}\sigma(p_1,p_2)} \; ,
\end{equation}
and
\begin{equation}
E_2(t,e) \; := \;
\int_{-\pi}^\pi \frac{d^2p}{(2\pi)^2} \;
\frac{2-2\cos(p_2 n)}{{p}^{2}} \;
\frac{{p}^{2} \rho^\prime(p_1,p_2)  }{
1 + \frac{e^2}{\pi}\sigma(p_1,p_2)} \; .
\end{equation}
Using the fact that $1 \geq (2-2\cos p_\mu)/p_\mu^2 > 1/5$ for $p_\mu
\in [ -\pi,\pi]$, one can estimate
\[
E_1(t,e) \; \leq \; \int_{-\pi}^\pi \frac{d^2p}{(2\pi)^2} \;
\frac{2-2\cos(p_2 t)}{{p}^2} \;
\frac{ \frac{e^2}{{p}^{2}} }{
1 + \frac{1}{\pi 25}\frac{e^2}{{p}{^2}}
+ \frac{e^2}{\pi}\sigma^\prime(p_1,p_2)}
\]
\[
= \; \int_{-\pi/e}^{\pi/e} \frac{d^2q}{(2\pi)^2} \;
\frac{2-2\cos(q_2 e t)}{{q}^{2}} \;
\frac{ 1 }{{q}^{2} + \frac{1}{\pi 25}
+ \frac{e^2}{\pi}\sigma^\prime(eq_1,eq_2) {q}^{2}} \; \]
\begin{equation}
\leq \;
\int_{-\infty}^{\infty} \frac{d^2q}{(2\pi)^2} \;
\frac{2-2\cos(q_2 e_c |{x}|)}{{q}^{2}} \;
\frac{ 1 }{ {q}^{2} + \frac{1}{\pi 25}} \; < \; \infty \; ,
\end{equation}
where the variable transformation ${q} = {p}/e$ was performed in
the second step and (52) was inserted in the last step. $E_2(t,e)$ is
even simpler shown to be bounded
\begin{equation}
E_2(t,e) \; \leq \; e^2 t^2 \; \int_{-\pi}^\pi \frac{d^2p}{(2\pi)^2} \;
\frac{p_2^2}{{p}^{2}} \; p^{2}  \rho^\prime(p_1,p_2)
\; < \; \infty \; .
\end{equation}
Thus the exponent $E(t,e) = E_1(t,e) + E_2(t,e)$ remains bounded and
the only factor that causes renormalization comes from the $1/t^2$
term in (41). The wave function renormalization $Z_\chi(e)$ for the
chiral densities can be chosen equal to $Z_{12}(e)$ and the continuum
limit of the two point function of the chiral densities is defined as
\begin{equation}
\lim_{{e \rightarrow 0 \atop et = const}} \; Z_\chi(e) \;
\langle \overline{\psi}(0,t) P_+ \psi(0,t) \;
\overline{\psi} (0,0) P_- \psi(0,0) \rangle \; \; \; \;
\mbox{where} \; \; \; \; Z_\chi(e) = \frac{e_c^2}{e^2} \; .
\end{equation}
This completes the discussion of the wave function renormalization.
The continuum charge $e_c$ entering $Z_{12}(e)$ and $Z_\chi(e)$ is obtained
from the rate at which the lattice coupling $e$ is driven to zero
(compare (52)).

\section{Concluding remarks}
It has been demonstrated that the Schwinger model with interpolated gauge
fields in combination with a continuum fermion determinant gives an
interesting effective lattice gauge theory. It was shown that the model
has a critical point where the continuum limit can be taken.
It reproduces the Schwinger model in the continuum. The
continuum model is well understood, and one knows that the model is
OS-positive from its bosonization to free fields \cite{lowenstein}.
However, usually
one wants to construct a continuum model from a lattice model at
the critical point and no independent approach to the continuum theory
is available. In particular one would like to prove OS-positivity already
for the lattice model, in order to have it granted for the
continuum limit. The standard proof for the Wilson formulation of lattice
gauge theories \cite{oster} makes use of an explicit
representation of the functional integral for the lattice fermions. This
possibility is lost when one considers the effective lattice gauge model
obtained from the continuum determinant. On the other hand,
the effective gauge field action is highly nonlocal, as can already
be seen from the Schwinger model. There the contribution $S_F$
from the fermion determinant (compare (15))
can be expressed in configuration space
using the Greens function for $- \triangle$ which is given
by $-1/4\pi \ln({x}^{2})$. Thus $S_F$ is a quadratic form
$\sum_{{n},{m}} F_{12}({n})F_{12}({m})
M({n},{m})$ for the lattice field strength
$F_{12}({n})$, with a kernel $M({n},{m})$
which behaves as $\ln( ({n} - {m})^{2})$
for large distances. This nonlocal behaviour even rules out the
strategy \cite{froh}. To prove OS-positivity one maybe has to
go back beyond the integration of the fermions. However, this unresolved
problem is a missing cornerstone of the approach using interpolated
gauge fields and a continuum definition of the fermion determinant.

{}From computing the chiral condensate in
the lattice model and comparing it to the N-flavor continuum Schwinger
model it was demonstrated that there is no fermion doubling which is
of course not surprising due to the usage of the continuum determinant.
Also the chiral properties of the continuum model are reproduced
already for finite correlation length.

Finally for the Schwinger model it was possible to entirely control
the continuum limit. The necessary wave function renormalization
constants for the field strength and the chiral densities were computed.
Together with the result for the correlation length this sheds light
on the problem of taking the continuum limit in an ab initio
lattice calculation \cite{lang}.
\vskip3mm
\noindent
{\bf Ackknowledgements :} \\
The author thanks Helmut Gausterer, Christian Lang, Erhard Seiler
and Larry Stuller for many fruitful discussions.

%
%-----------------------------------------------------------
%

\end{document}